\documentclass[11pt]{article}
\usepackage{sint,macros_ALPHA}
\usepackage{amssymb}
\usepackage[dvips]{epsfig}
\begin{document}
\begin{titlepage}

\begin{flushright}
DESY 99-024 \\
hep-lat/9903016
\end{flushright}

\vskip 1 cm
\begin{center}
{\Large\bf 
Scaling investigation of renormalized correlation functions\\[0.5ex]
in $\Or(\fat{a})$ improved quenched lattice QCD
}
\end{center}
\vskip 1 cm
\vbox{
\centerline{
\epsfxsize=2.5 true cm
\epsfbox{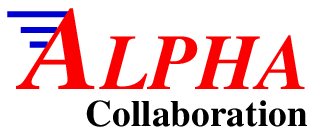}}
}
\vskip 1 cm
\begin{center}
{\large
Jochen Heitger
}
\vskip 1.0cm
Deutsches Elektronen-Synchrotron, DESY Zeuthen\\
Platanenallee 6, D-15738 Zeuthen, Germany
\vskip 1.5cm
{\bf Abstract}
\vskip 0.7ex
\end{center}
We present a scaling investigation of some correlation functions in
$\Or(a)$ improved quenched lattice QCD.
In particular, as one observable the renormalized PCAC quark mass is
considered.
Others are constructed such that they become the vector meson mass and
the pseudoscalar meson decay constant when the volume is large.
For the present discussion we remain in intermediate volume,
$(0.75^3\times1.5)\,\,\mbox{fm}^4$ with Schr\"odinger functional boundary
conditions.
By fixing the `pion mass' and the spatial lattice size in units of the
hadronic scale $r_0$, we simulated four lattices with resolutions ranging
from 0.1 fm to 0.05 fm and performed the extrapolation to the continuum
limit.
The maximal scaling violation found in the improved theory is a
$\sim$ 6 \% effect at $a\simeq 0.1\,\,\mbox{fm}$. \vfill

\begin{flushleft}
DESY 99-024 \\
March 1999
\end{flushleft} 

\eject
\vfill
\eject
\end{titlepage}

\section{Introduction}
\label{Intro}
One of the major drawbacks in the standard formulation of lattice
QCD, induced by the Wilson-Dirac operator violating chiral symmetry
at the scale of the cutoff \cite{WI:bochi,WI:maiani}, reflects among
others in the fact that the quark masses are not protected from additive
renormalizations and that the leading lattice effects in physical matrix
elements and amplitudes are proportional to the lattice spacing $a$
rather than being of $\Or(a^2)$.

Nowadays, a systematic approach based on the Symanzik improvement
programme \cite{impr:sym1,impr:sym2} is well established to permit a
removal of these discretization errors of $\Or(a)$ in lattice QCD with
confidence.
It has been elaborated for on-shell quantities in
refs.~\cite{luscher:1985zq,impr:onshell} and adds appropriate
higher-dimensional operators to the action and fields in order to
compensate for any correction terms at $\Or(a)$.

Within this framework a mostly non-perturbative $\Or(a)$ improvement of
the action and the quark currents as well as their renormalization has
been achieved in the quenched case, where quark loops are ignored.
The basic idea for the practical implementation of on-shell improvement
to enable a numerical computation of these improvement coefficients
--- pioneered by the ALPHA collaboration --- is to exploit chiral
symmetry restoration and certain chiral Ward identities from Euclidean
current algebra relations on the lattice at $\Or(a)$
\cite{impr:pap1,tsuk97:rainer}.
Most prominently, demanding the PCAC relation $\Pmu\amu=2m\p$ between
the isovector axial current $\amu$ and pseudoscalar density $\p$ to hold
as a renormalized operator identity, imposes a sensible improvement
condition in this respect.
As a result, the values of the improvement coefficients $\csw$, $\ca$
and $\cv$, which are required to completely eliminate the corresponding
$\Or(a)$ effects, are known for $\beta=6/g_0^2\ge6.0$, $g_0$ being the
bare gauge coupling \cite{impr:pap3,impr:pap4,lat97:marco}.

Therefore, one is now interested in the quality of scaling behaviour and
the size of its possible violation.
During the last two years it was reported in this context
\cite{impr:QCDSF_lett,impr:QCDSF,lat97:hartmut} that at
$a\simeq 0.1\,\,\mbox{fm}$ the residual $\Or(a^2)$ lattice artifacts
may be still fairly large e.g.~for the kaon decay constant $f_Kr_0$
($\sim$ 10 \%), while they are very small already for other quantities
like the rho meson mass $\mrh/\sqrt{\sigma}$ ($\sim$ 2 \%) 
\cite{lat97:hartmut,impr:SCRI}.
Thus, restricting to an intermediate physical volume, the present scaling
tests are intended to examine the impact of full $\Or(a)$ improvement in
quenched lattice QCD thoroughly and with high accuracy for different
observables.
Some of these are designed to coincide with phenomenologically relevant
observables in the limit of large physical volumes.
A preliminary status of this work has been briefly surveyed recently in
ref.~\cite{lat98:jochen}.

We wish to point out that the present investigation must be
distinguished from similar studies like
\cite{impr:QCDSF,impr:roma2_2,impr:roma1_2} for an important reason.
Namely, since we address directly the scaling behaviour of correlation
functions in finite volume, we do not have to rely on the asymptotic
behaviour of (ratios of) usual timeslice correlation functions to
extract, for instance, hadron masses or decay constants; these more
conventional techniques often reveal to be genuinely affected by
systematic errors difficult to control.
Actually, an avoidance of such intrinsic uncertainties is supplied in
part by the Schr\"odinger functional: its finite-volume fermionic
correlation functions are scaling quantities at any fixed distance in
time, if the renormalization factors of the quark fields at the
boundaries are properly divided out.
Beyond that, they decay very slowly for small time separations, allow to
gain a good numerical precision and hence offer the appealing
possibility to probe the theory for $\Or(a)$ improvement without unwanted
additional sources of errors.
Finally, for sufficiently large volumes the correlators can be shown to
embody standard hadronic masses and matrix elements \cite{msbar:pap2}.
These aspects provide the real advantages of our method.
The prize to pay for it is, however, that within the Schr\"odinger
functional formulation there exist two further improvement coefficients
multiplying the boundary counterterms.
These are perturbatively known only and as a consequence, one can in
principle not exclude that the theory is to some extent still
contaminated with uncancelled $\Or(a)$ contributions. 
That this fear can be dismissed in practice has recently been
demonstrated for the renormalization group invariant running quark
mass \cite{msbar:pap1}, but we will make sure of it in the present
context too.

The paper is organized as follows.
In section~\ref{CFs} we introduce the fermionic correlation functions
under study and sketch how spectral observables can be constructed from
them in the Schr\"odinger functional scheme.
After a short account on the numerical simulations, section~\ref{ScalT}
contains a detailed description of the scaling test and the careful
evaluation of the data.
Section~\ref{Res} gives the results. Here, also the question will be
answered whether an improvement condition, chosen to fix a certain
improvement coefficient non-perturbatively, is unambiguous in the sense
that it automatically implies appreciably small scaling violations of
$\Or(a^2)$ in other quantities not related to this specific condition.
We conclude with a discussion in section~\ref{DiscOut}.

\section{Correlation functions and hadronic observables}
\label{CFs}
The basic framework for our lattice setup is the QCD Schr\"odinger
functional (SF), whose concepts and characteristic features have been
published in much detail in refs.~\cite{SF:LNWW,SF:stefan1,SF:stefan2};
consult also \cite{schlad:rainer,reviews:leshouches} for comprehensive
overviews on the subject.

It is defined as the partition function of QCD in a cylindrically shaped
space-time manifold of extension $L^3\times T$ with periodic boundary
conditions in the space directions and (in general inhomogeneous)
Dirichlet boundary conditions at times $x_0=0$ and $x_0=T$. 
This means in the case of the gluons to require the spatial gauge field
components at the boundaries to satisfy $A_k(x)|_{x_0=0}=C_k(\bx)$ and
$A_k(x)|_{x_0=T}=C_k'(\bx)$, where $C_k$ and $C_k'$ are some prescribed
smooth classical (chromo-electric) gauge potentials, and a similar 
assignment is imposed on the quark fields as well.
One of the bold advantages of such a choice is that it ensures frequency
gaps for the gluon and quark fields, and thereby numerical simulations at
vanishing quark masses become tractable.
As in most of the other applications of the QCD SF, we assume the
special choice of homogeneous boundary conditions from now on:
$C_k=C_k'=0$ for the spatial components of the gauge potentials and
vanishing fermion boundary fields.
\subsection{Correlation functions in the Schr\"odinger functional} 
\label{CFinSF}
Although the definitions of fermionic correlation functions within the SF
already appeared in the literature
\cite{impr:pap1,impr:pap3,impr:pap4,lat97:marco}, let us recall them here
and collect the essential properties and formulae in order to make the
paper self-contained.

If $\zeta$ and $\zeb$ denote `boundary quark and antiquark fields' at 
Euclidean time $x_0=0$ and primed symbols the corresponding objects at
$x_0=T$ \cite{impr:pap1}, one builds up the boundary field products 
\be
\obi=a^6\sum_{\bys,\bzs}\zeb(\by)\gfv\,\gen\,\zeta(\bz)\,,\quad
\obf=a^6\sum_{\bus,\bvs}\zeb'(\bu)\gfv\,\gen\,\zeta'(\bv)
\label{bound_o}
\ee
and analogously,
\be
\qbi=a^6\sum_{\bys,\bzs}\zeb(\by)\gamma_k\,\gen\,\zeta(\bz)\,,\quad
\qbf=a^6\sum_{\bus,\bvs}\zeb'(\bu)\gamma_k\,\gen\,\zeta'(\bv)\,,
\label{bound_q}
\ee
where $\tau^a$, $a=1,2,3$, are the Pauli-matrices acting on the first two
flavour components of the quark fields $\psi$.
In the operator language of quantum field theory they create initial and
final quark-antiquark states with zero momenta, respectively, and
transform according to the vector representation of the exact isospin
symmetry.
For the axial-vector current $\amu$ and the pseudoscalar density $\p$ we
use the local expressions
\be
\amu(x)=\bpsi(x)\gmu\gfv\,\gen\,\psi(x)\,,\quad
\p(x)=\bpsi(x)\gfv\,\gen\,\psi(x)\,,
\label{amu_and_p}
\ee
while the vector current $\vmu$ and the anti-symmetric tensor field
$\tmunu$ read
\be
\vmu(x)=\bpsi(x)\gmu\,\gen\,\psi(x)\,,\quad
\tmunu(x)=i\,\bpsi(x)\smunu\,\gen\,\psi(x)\,.
\label{vmu_and_tmunu}
\ee
Now we consider correlation functions on the lattice in the SF.
By inserting the preceding densities at some inner point $x$ of the SF
cylinder (with support on the hypersurface at $x_0$) between the
appropriate external quark-antiquark states, one introduces the
expectation values
\bea
\fa(x_0)
& = &
-a^6\sum_{\bys,\bzs}\frac{1}{3}\,
\mvl{\az(x)\,\zeb(\by)\gfv\,\gen\,\zeta(\bz)}\,\,\,=\,\,\,
-\frac{1}{3}\,\mvl{\az(x)\,\obi}
\label{cor_fa}\\
\fp(x_0)
& = &
-a^6\sum_{\bys,\bzs}\frac{1}{3}\,
\mvl{\p(x)\,\zeb(\by)\gfv\,\gen\,\zeta(\bz)}\,\,\,=\,\,\,
-\frac{1}{3}\,\mvl{\p(x)\,\obi}\,,
\label{cor_fp}
\eea
analogously,
\bea
\kv(x_0)
& = &-\frac{1}{9}\,\mvl{V_k^a(x)\,\qbi}
\label{cor_kv}\\
\kt(x_0)
& = &-\frac{1}{9}\,\mvl{\tkz(x)\,\qbi}\,,
\label{kor_kt}
\eea
and the boundary-boundary correlation function
\be
f_1=-\frac{1}{3L^6}\,\mvl{\obf\obi}\,.
\label{cor_f1}
\ee
Gauge invariant correlators of this type have already been used to study
the conservation of currents on the lattice and to deduce suitable
improvement and normalization conditions in lattice QCD in order to
calculate the corresponding coefficients and constants non-perturbatively
by numerical simulations \cite{impr:pap3,impr:pap4,lat97:marco}.
One of them, $f_1$, will be utilized later to cancel the multiplicative
renormalization of the boundary quark fields $\zeta,\ldots,\zeb'$.
Besides on the SF characteristic kinematical variables \cite{impr:pap1},
the correlation functions depend on the bare parameters $g_0$, $m_0$ and
the improvement coefficient $\csw=\csw(g_0)$ in the fermionic part of the
lattice action, but not on the spatial coordinates of $x$ owing to
translation invariance.
There is also a dependence on the improvement coefficients $\ct$ and
$\ctt$, which account for specific boundary $\Or(a)$ counterterms arising
in the SF approach \cite{impr:pap1}.

After contracting the quark fields, all the bare and unimproved
correlation functions have the general structure
$\hX(x_0)\propto\mvl{\Tr\{H^+(x)\dts{\Gamma}{X}H(x)\}}$, $h=f,k$,
where the respective insertions are
$\dts{\Gamma}{X}\in\{-\gamma_0,1,\gamma_k,-\sigma_{k0}\}$, X=A,P,V,T;
the trace extends over colour, Dirac and (in principle as well) flavour
indices, and the matrix $H$ is the quark propagator from the boundary at
$x_0=0$ to the point $x$ in the interior of the space-time volume
\cite{impr:pap3}.
In the quenched approximation the expectation values are understood to be
taken as path integral averages in the pure gauge theory.

We should mention that also the primed correlation functions $\hX'$, which
are connected to $\hX$ through a time reflection and vice versa, become
relevant.
Those are in the case of eqs.~(\ref{cor_fa}) and (\ref{cor_fp})
\be
\fa'(T-x_0)=+\frac{1}{3}\,\mvl{\az(x)\,\obf}\,,\quad
\fp'(T-x_0)=-\frac{1}{3}\,\mvl{\p(x)\,\obf}\,,
\label{cor_fx_refl}
\ee
and similar relations apply to $\kv,\kv'$ and $\kt,\kt'$.
Obviously, in $\hX'$ the currents and densities are probed by the boundary
quark fields at $x_0=T$ instead, and the argument $T-x_0$ indicates that
they fall off with this distance.
For vanishing gauge fields $C$ and $C'$ at the boundaries, our correlation
functions possess the useful time reflection invariance
$\hX(x_0)=\hX'(x_0)$.
This allows to sum them up accordingly, and averaging over the spatial
components helps to reduce the statistical noise in the Monte Carlo
simulation further.

On-shell improvement at $\Or(a)$ for the axial and vector currents is
achieved by adding the derivatives of the pseudoscalar density and the
tensor current as the suitable $\Or(a)$ counterterms,
\bea
\amui(x)
& \equiv &\amu(x)+a\ca\Smu\p(x)
\label{cur_a_i}\\
\vmui(x)
& \equiv &\vmu(x)+a\cv\Snu\tmunu(x)\,,
\label{cur_v_i}
\eea
where the improvement coefficients $\ca$ and $\cv$ are determined by the
demand to cancel the $\Or(a)$ errors in lattice Ward identities, emerging
from a mixing with higher-dimensional operators with the same quantum
numbers \cite{impr:pap4,lat97:marco}.
Then the corresponding improved fermionic correlation functions are given
by:
\bea
\fai(x_0)
& = &\fa(x_0)+a\ca\Sz\fp(x_0)
\label{cor_fa_i}\\
\kvi(x_0)
& = &\kv(x_0)+a\cv\Sz\kt(x_0)\,.
\label{cor_kv_i}
\eea
The lattice derivative $\Smu\equiv\half\,(\Pmu+\Pmu^*)$ is the symmetrized
combination of the usual forward and backward difference operators $\Pmu$
and $\Pmu^*$, acting as
\[
\Pmu f(x)=\frac{f(x+a\hat{\mu})-f(x)}{a}\,,\quad
\Pmu^* f(x)=\frac{f(x)-f(x-a\hat{\mu})}{a}\,.
\]
Herewith we are already in the position to write down the unrenormalized
PCAC quark mass as a function of the timeslice location $a\le x_0\le T-a$:
\be
m(x_0)=\frac{\Sz\fa(x_0)+a\ca\Pz^*\Pz\fp(x_0)}{2\fp(x_0)}\,.
\label{mcurr_x0}
\ee
For the properly chosen value of $\ca=\ca(g_0)$ at given gauge coupling
$g_0$ and any hopping parameter $\kappa$ it is defined by obeying the
PCAC relation (for two degenerate quark flavours) up to cutoff effects of
$\Or(a^2)$,
\be
\Smu[\,\amui(x)\,]=2m\p(x)+\Or(a^2)\,,
\label{rel_pcac}
\ee
which more rigorously must be looked at as a renormalized operator
identity 
\[
\Smu\mvl{\amur(x)\ob}=2\mr\mvl{\pr(x)\ob}+\Or(a^2)
\]
in terms of some arbitrary renormalized on-shell $\Or(a)$ improved field
$\ob$ localized in a region not containing $x$.
\subsection{Construction of renormalized observables}
\label{ConObs}
Now we want to introduce the renormalized correlation functions and
design the scaling combinations of them, which will be studied
numerically in the next section.

The QCD SF serves as a particular intermediate finite-volume
renormalization scheme which is, however, not necessarily related to a
special regularization \cite{impr:pap1,schlad:rainer,reviews:leshouches}. 
Here, the SF is employed as a mass-independent renormalization scheme,
while the ratio $T/L$ is assumed to be kept fixed to a certain value. 
The freedom in choosing the boundary fields $C$ and $C'$ (as well as the
boundary conditions on the quark fields specified by angles
$\theta_{\mu}$, cf.~subsection~\ref{MCsim}) different from zero are left
for other applications, see e.g.~\cite{impr:pap3,msbar:pap1,alpha:SU3}.
Moreover, the SF respects $\Or(a)$ improvement after adding the $\Or(a)$
counterterms $\propto\ct,\ctt$ so that by attaching additive and
multiplicative renormalization constants, the quantities
\bea
\amur(x)
& = &\za(1+\ba a\mq)\amui(x)
\label{cur_axl_r}\\
\vmur(x)
& = &\zv(1+\bV a\mq)\vmui(x)
\label{cur_vec_r}\\
\pr(x)
& = &\zp(1+\bP a\mq)\p(x)
\label{den_psc_r}
\eea
induce the renormalized and improved correlation functions
\bea
\far(x_0)
& = &\za(1+\ba a\mq)\zzet^2(1+\bzet a\mq)^2\fai(x_0)
\label{cor_fa_r}\\
\kvr(x_0)
& = &\zv(1+\bV a\mq)\zzet^2(1+\bzet a\mq)^2\kvi(x_0)
\label{cor_kv_r}\\
\fpr(x_0)
& = &\zp(1+\bP a\mq)\zzet^2(1+\bzet a\mq)^2\fp(x_0)
\label{cor_fp_r}\\
{\fr}
& = &\zzet^4(1+\bzet a\mq)^4 f_1\,.
\eea
The constants $\zzet$ and $\bzet$ (the former being scale dependent
\cite{impr:pap1}) have to be attributed to the boundary values of the
quark and antiquark fields appearing in the products (\ref{bound_o}) and
(\ref{bound_q}) in the renormalized theory. 
In a mass-independent renormalization scheme the underlying $\Or(a)$
counterterm enters as
\[
\zer(\bx)=\zzet(1+\bzet a\mq)\zeta(\bx)
\]
and similarly for the antiquark field $\zeb$, giving
$\obir=\zzet^2(1+\bzet a\mq)^2\obi$ for instance.

Because the renormalization does not distinguish between different
flavours in a mass-independent scheme, the knowledge of the $Z$--factors
suffices to link the lattice theory at finite cutoff to the renormalized
continuum theory. 
Here, all normalization conditions to fix and determine the
renormalization constants $\zX$ and $\bX$, X=A,V,P, were imposed on
appropriate matrix elements at \emph{zero quark mass}, which is safe
within the SF scheme as the finite lattice extent $L$ provides the
before-mentioned natural infrared cutoff for the theory
\cite{impr:pap3,impr:pap4}.
Because of the zero quark mass condition they are functions of $g_0$ only
and \emph{not} of the subtracted bare quark mass, which equals
\be
a\mq=am_0-a\mc(g_0)=
\frac{1}{2}\,\left(\frac{1}{\kappa}-\frac{1}{\kapc}\right)
\label{m_quark}
\ee
and vanishes along a critical line $m_0=\mc(g_0)$ in the plane of bare
parameters, implicitly defining the critical hopping parameter $\kappa_c$.
Any remaining corrections of $\Or(a\mq)$ are supposed to be cancelled by
adjusting the $\bX$ alone.

Now we can pass to the set of observables we have constructed for the
present study.
We start with the renormalized PCAC (current) quark mass in the SF scheme,
which in view of eqs.~(\ref{mcurr_x0}), (\ref{cur_axl_r}) and
(\ref{den_psc_r}) may be defined as
\be
\mbar=\frac{\za}{\zp(L)}\,m\left({\T \frac{T}{2}}\right)
\label{mpcac}
\ee
by multiplying the proper renormalization constants $\za$ and
$\zp(L)$, the latter assumed to be taken at some renormalization scale
$\mu=1/L$ \cite{msbar:pap1,lat97:martin}.
Strictly speaking, the ratio of the additive renormalization factors
$1+\ba a\mq$ and $1+\bP a\mq$ would have been to be accounted for as well,
but it turns out perturbatively \cite{impr:pap5} and 
non-perturbatively \cite{impr:roma2_1} that $(\ba-\bP)a\mq$ is numerically
quite small at the interesting values of the bare gauge coupling and the
hopping parameter.
Hence we neglect it, here.

Beyond that, we compose the following (time dependent) combinations of
renormalized and improved fermionic correlation functions.
Firstly, the logarithmic time derivatives 
\bea
\mups(x_0) 
& = &
\frac{\Sz\fpr(x_0)}{\fpr(x_0)}
\,\,\,=\,\,\,\frac{\Sz\fp(x_0)}{\fp(x_0)}
\label{mpi_x0}\\
\muV(x_0)
& = &
\frac{\Sz\kvr(x_0)}{\kvr(x_0)}
\,\,\,=\,\,\,\frac{\Sz\kvi(x_0)}{\kvi(x_0)}
\label{mrho_x0}
\eea
of the respective SF correlation functions in the pseudoscalar and vector
meson channel; they deviate from the ordinary definition of effective
masses ($\sim\partial_0\ln\hX$) by terms of $\Or(a^2)$.
Secondly, we will consider the ratios
\bea
\etps(x_0)         
& = &
\dts{C}{PS}\,\frac{\far(x_0)}{\sqrt{\fr}}\nonumber\\
& = &
\dts{C}{PS}\,\frac{\za(1+\ba a\mq)\fai(x_0)}{\sqrt{f_1}}\,,\quad
\dts{C}{PS}=\frac{2}{\sqrt{L^3\mups(\frac{T}{2})}}
\label{fpi_x0}\\
\etV(x_0)
& = &
\dts{C}{V}\,\frac{\kvr(x_0)}{\sqrt{\fr}}\nonumber\\
& = &
\dts{C}{V}\,\frac{\zv(1+\bV a\mq)\kvi(x_0)}{\sqrt{f_1}}\,,\quad
\dts{C}{V}=\frac{2}{\sqrt{L^3\,[\,\muV(\frac{T}{2})\,]^3}}\,.
\label{frho_x0}
\eea
Since through the division by $\sqrt{\fr}$ the renormalization factors of
the boundary quark fields, $\zzet(1+\bzet a\mq)$, drop out, it is ensured
that $\etps$ and $\etV$ (at fixed argument $x_0$) exhibit scaling and have
a well-defined continuum limit.
We note in passing that alternatively the correlation function $\fai$ also
could have been used instead of $\fp$ to define a `local mass' in the
pseudoscalar channel.
But as $\fai$ amounts to somewhat larger statistical errors in the time
derivatives, we here preferred $\fp$ for the purpose of $\mups(x_0)$.

In the present scaling test we fix a definite temporal separation from the
boundaries, $x_0=T/2$.
Thus one arrives at the objects
\bea
\mbox{pseudoscalar channel}:
& \quad &
\mups\left({\T \frac{T}{2}}\right)\quad
\etps\left({\T \frac{T}{2}}\right)
\label{mpi_and_fpi}\\
\mbox{vector channel}:
& \quad &
\muV\left({\T \frac{T}{2}}\right)\quad
\etV\left({\T \frac{T}{2}}\right)\,.
\label{mrho_and_frho}
\eea
Their continuum limits are expected to be approached like 
$\muX(\frac{T}{2})+\Or(a^2)$ and $\etX(\frac{T}{2})+\Or(a^2)$, X=PS,V,
in the $\Or(a)$ improved theory.
The choice $x_0=T/2$ is motivated by the fact that in the SF scheme cutoff
effects are generically larger when sitting closer to the boundaries.
It is possible to construct other quantities in a similar way, but we
consider the foregoing ones as reasonably representative.

Let us emphasize, however, a final point.
Adopting the quantum mechanical representation of the field operators
associated to (\ref{amu_and_p}) and (\ref{vmu_and_tmunu}), it can be shown
in the transfer matrix formalism that asymptotically for large Euclidean
times the quantities in eqs.~(\ref{mpi_and_fpi}) and (\ref{mrho_and_frho})
become the pseudoscalar and vector meson masses ($\mups,\muV$) as well as
the pseudoscalar decay constant ($\etps$).
For instance, the proportionality constant $\dts{C}{PS}$ in (\ref{fpi_x0})
is such that this ratio turns, as the temporal lattice extent goes to
infinity ($x_0,T-x_0\rightarrow\infty$), into a familiar matrix element,
which complies with the standard definition of the pion decay constants in
continuum QCD:
\[
\za\ketbra{0}{\bpsi(x)\gamma_0\gfv\,\gen\,\psi(x)}{\pi^a(\bn)}=\mpi\fpi\,.
\]
More formally, the observables $\cl{O}$ just introduced should be
regarded as functions $\cl{O}(T/L,x_0/L,L/r_0,a/r_0)$, and in the spirit
of the above a physically meaningful situation is realized if, as the
spatial volume $(L/a)^3$ tends to infinity, $x_0\gg r_0$ and
$T-x_0\gg r_0$ are valid for some typical hadronic radius of 
$r_0\simeq 0.5$ fm.
Masses and matrix elements of interest in hadron phenomenology can then be
extracted \cite{msbar:pap2}.

\section{Scaling tests}
\label{ScalT}
Before going into the details of the investigation, one has to keep in
mind that here the actually taken lattice volumes in physical units are
only of intermediate magnitude.
Therefore, any of the following results are prevented from resembling the
large volume limit, where our observables were argued to receive a really
physical meaning.
Instead of this, most emphasis is on the scaling properties of the theory,
which should not depend on the specific choice of the lattice size and
the SF characteristic boundary conditions.
Nevertheless, we refer from now on to the (first three of the) quantities
in eqs.~(\ref{mpi_and_fpi}) and (\ref{mrho_and_frho}) as the `pion mass',
its decay constant and the `rho meson mass' and assign the common symbols
$\mpi$, $\fpi$ and $\mrh$ to them in the obvious manner.
We also denote $\etVb\equiv\etV(\frac{T}{2})$ in the vector meson channel.

The advantage of working at finite quark mass in a direct test of
improvement should be stressed explicitly.
Namely, this avoids any extrapolations and evades the potential problems
with so-called exceptional configurations one runs into, when the parameter
region of zero quark mass is attempted to be reached
\cite{impr:pap3,exconf:FNAL_1}.
\subsection{Monte Carlo simulation}
\label{MCsim}
The cost of a quenched QCD simulation is always governed by the
computation of fermion propagators required for the correlation functions
to be covered.
This involves the action of the Wilson-Dirac operator on quark fields
$\hf{\psi}$, which for the $\Or(a)$ improved theory in the framework of the
SF is conveniently decomposed as $D+\delta D+m_0$ with
\be
(D+m_0)\hf{\psi}\equiv\frac{1}{2\kappa}\,M\hf{\psi}\,,\quad
\kappa=\frac{1}{8+2m_0}
\label{w_dir}
\ee
\bea
M\hf{\psi}
& = &
\hf{\psi}-\kappa\sum_{\mu=0}^{3}\Big\{
\lambda_{\mu}\gf{U}{\mu}(1-\gmu)\hfup{\psi}{\mu}\nonumber\\
&   &\hspace{2.45cm}
+\lambda_{\mu}^*\gfdn{U}{\mu}{\mu}^+(1+\gmu)\hfdn{\psi}{\mu}\Big\}\,,
\label{w_dir_act}
\eea
where, as usual, the fermionic degrees of freedom $\hf{\psi}$ live on the
sites $x$ of the lattice, and $\gf{U}{\mu}$ denotes the SU(3)--valued gauge
links in lattice direction $x+a\hat{\mu}$, $\mu=0,\dots,3$.
The factors $\lambda_{\mu}=\Exp^{\,ia\theta_{\mu}/L}$, $\theta_0\equiv0$,
give rise to a modified covariant derivative equivalent to demanding
spatial periodicity of the quark field up to a phase $\Exp^{\,i\theta_k}$;
for our purposes $\theta_k$, $k=1,2,3$, was set to zero
throughout.\footnote{
In general, $\theta_k$ can serve as an additional kinematical variable to
formulate proper improvement and normalization conditions,
see for instance \cite{impr:pap3,msbar:pap1}.}
The local $\Or(a)$ counterterm $\delta D=\delta\dts{D}{v}+\delta\dts{D}{b}$
now consists of two contributions, namely the Sheikholeslami-Wohlert
clover term \cite{impr:SW}
\be
\delta\dts{D}{v}\hf{\psi}=
\csw\,\frac{i}{4}\,a\smunu\pl{F}{\mu\nu}\hf{\psi}
\label{w_dir_clover}
\ee
and a term
\bea
\delta\dts{D}{b}\hf{\psi}
& = &
(\ctt-1)\Big\{\delta_{x_0/a,1}\big[\,
\hf{\psi}-\gfdn{U}{0}{0}^+P_+\hfdn{\psi}{0}\,\big]\nonumber\\
&   &\hspace{1.6cm}
+\delta_{x_0/a,T-1}\big[\,\hf{\psi}-\gf{U}{0}P_-\hfup{\psi}{0}\,\big]\Big\}
\label{w_dir_bound}
\eea
with 
\[
P_+\hf{\psi}\Big|_{x_0=0}=0\,,\quad
P_-\hf{\psi}\Big|_{x_0=T}=0\,,\quad
P_{\pm}\equiv\half\,(1\pm\gamma_0)\,,
\]
which is specific for the SF type of boundary conditions in our setup.
The improvement coefficient $\ctt$ and a further one, $\ct$ which enters
the calculation as well but is independent of the local composite operators
containing quark fields and thus not written down here, are set to their
one-loop perturbative values.\footnote{
Recently, the coefficient $\ct$ has also been computed up to two-loop order
of perturbation theory in the quenched case \cite{pert:2loop_2}.}

Since the technicalities of the Monte Carlo simulations are identical to
those already detailed in ref.~\cite{impr:pap3}, it is not necessary to
repeat them here in full length.
Our data were taken on the APE-100 massively parallel computers with
128 -- 512 nodes at DESY Zeuthen, whose topology also allows to simulate
independent replica of the system at the same time in the case of smaller
lattice volumes (sets A -- C below).
The gauge field ensembles were generated by a standard hybrid
overrelaxation algorithm, where each iteration consists of one heatbath
step followed by several microcanonical reflection steps (typically
$\dts{N}{OR}=L/2a+1$), and the correlation functions have been evaluated by
averaging over sequential gauge field configurations separated by 50
iterations.
To solve the system of linear equations belonging to the boundary value
problem of the Dirac operator within the measurements of the correlators,
the BiCGStab algorithm with even-odd preconditioning was used as inverter.
Finally, a single-elimination jackknife procedure was applied to estimate
the statistical errors of all the secondary quantities, because the data
stemming from the same configurations must be considered as strongly
correlated.
By dividing the full ensemble of measurements into bins we also checked for
the statistical independence of our data samples.
\subsection{Method and numerical analysis}
\label{Meth}
For the analysis we use $\csw$, $\cX$, $\zX$, X=A,V,P, and $\bV$
non-perturbatively determined in
\cite{impr:pap3,impr:pap4,lat97:marco,msbar:pap1} for $\beta\ge6.0$, while
$\ba$ and $\bP$ are taken from one-loop perturbation theory
\cite{impr:pap1,tsuk97:rainer,impr:pap5}.
If available, we always adopted the rational formulas for the former with
overall, i.e.~statistical and systematic, uncertainties stated in the
references.
As opposed to $\za$ and $\zv$ (and as already anticipated below
eq.~(\ref{mpcac}) in subsection~\ref{ConObs}), the normalization constant
of the pseudoscalar density, $\zp=\zp(L)$, acquires a scale dependence
through its renormalization.
The scale evolution of $\zp$, which due to its definition at the point of
vanishing PCAC quark mass (but in the absence of exact chiral symmetry at
finite $a$) is unique only up to $\Or(a^2)$ errors, has been recently
computed non-perturbatively in \cite{msbar:pap1}.
In the SF the appropriate renormalization scale is
$\mu=1/L\equiv 1/2\lmax$, with $\lmax/r_0=0.718$ \cite{pot:r0_ALPHA}, and
we take over the needed numbers for $\zp$ from the last but one reference. 
In the case of $\cv$ and the critical hopping parameter $\kappa_c$ in
eq.~(\ref{m_quark}), where no closed expressions are recommended yet, we
adapted the numbers at discrete values of the gauge coupling from
refs.~\cite{impr:pap3,lat97:marco} to our simulated $\beta$--values by
linear interpolation.

The strategy was then to keep a finite physical volume and the quark mass
fixed by prescribing the geometry $T/L=2$ and two further renormalization
conditions, which we decided to chose as
\be
\mpi L=2.0 \quad\mbox{and}\quad
\frac{L}{r_0}=1.49
\label{cond_LCP}
\ee
for the `pseudoscalar meson (pion) mass' and the spatial lattice size,
respectively.
The first condition on $\mpi$ can be, at least approximately\footnote{
In our actual simulations we have some mismatch in $\mpi L$ between the
individual points in parameter space, which will be discussed later.},
satisfied by a careful tuning of the hopping parameter 
($a^2\mpi^2\sim\kappa$).
Here the reference scale is expressed by the hadronic radius $r_0$ defined
in \cite{pot:r0} through the force between static quarks to yield the
phenomenologically motivated value of $r_0\simeq 0.5\,\,\mbox{fm}$.
Using the latest results on the hadronic scale $r_0/a$
in ref.~\cite{pot:r0_ALPHA}, quoted there as
\be
\ln\left(\frac{a}{r_0}\right)=
-1.6805-1.7139(\beta-6)+0.8155(\beta-6)^2-0.6667(\beta-6)^3\,,
\label{r0_formula}
\ee
one can solve numerically for $\beta$ after inserting $a/r_0=1.49a/L$ to
find the desired pairs $(L/a,\beta)$ in order to fulfill the second
condition in eq.~(\ref{cond_LCP}) within errors.
In practice, the particular value $L/r_0=1.49$ was determined by the 
initial simulations at $\beta=6.0$ on lattices with spatial size $L/a=8$,
and the larger lattices were adjusted thereafter.
The simulation parameters and some results are compiled in
table~\ref{ParTab}.
%
\begin{table}[htb]
\begin{center}
\begin{tabular}{|c|cccc|ccc|}
\hline
  set & $L/a$ & $\beta$ & $\kappa$ & $\dts{N}{meas}$ & $L/r_0$ 
& $a\mpi$ & $\mpi L$ \\
\hline
  A   & 8  & 6.0  & 0.13458 & 12800 & 1.490(6) & 0.2505(11) & 2.004(9)  \\
  B   & 10 & 6.14 & 0.13538 & 3840  & 1.486(7) & 0.1945(14) & 1.946(14) \\
  C   & 12 & 6.26 & 0.13546 & 2560  & 1.495(7) & 0.1709(13) & 2.050(16) \\
  D   & 16 & 6.48 & 0.13541 & 3000  & 1.468(8) & 0.1244(9)  & 1.991(15) \\
\hline
\end{tabular}
\caption[t_param]{\label{ParTab} \sl
                  Simulation points and its statistics $\dts{N}{meas}$
                  denoting the number of gauge field configurations, on
                  which the fermionic correlation functions were computed.
                  $L/r_0$ and $\mpi L$ are the quantities chosen to fix
                  renormalization conditions for the LCP studied.}
\end{center}
\end{table}
%
These settings give an intermediate volume of
$(0.75^3\times1.5)\,\,\mbox{fm}^4$, and one moves on a line of constant
physics (LCP) in bare lattice parameter space with lattice resolutions
ranging from 0.1 fm to 0.05 fm.

As an important prerequisite for the reliability of the scaling test we
had, of course, to estimate the dependence of our results on the SF
specific (and solely perturbatively known) improvement coefficients of the
boundary counterterms $\ct$ and $\ctt$, which with respect to full
non-perturbative $\Or(a)$ improvement represent the only imperfectly known
input parameters for the simulation.
Otherwise a complete suppression of errors linear in $a$ would not be
guaranteed, and continuum limit extrapolations with an $\Or(a^2)$ term as
the dominant scaling violation is a priori not justified.
To this end we verified by an artificial variation of the one-loop
coefficients in the expansions \cite{alpha:SU3,impr:pap2}
\bea
\uts{\ct}{1--loop}  & = &1-0.089g_0^2+\Or(g_0^4) 
\label{ct_1loop}\\
\uts{\ctt}{1--loop} & = &1-0.018g_0^2+\Or(g_0^4)  
\label{cttil_1loop}
\eea
by a factor 2 for $\uts{\ct}{1--loop}$ and by a factor 10 for
$\uts{\ctt}{1--loop}$ that at unchanged renormalization conditions
(\ref{cond_LCP}) their influence on the level of numerical precision in
our data is small enough to be neglected: it typically came out to be
below 1 \% for $a\fpi$, below 2 \% for $\etVb$, and nearly not visible for
$am$ and $a\mrh$, at parameters corresponding to simulation point A
($T/a=16$).
It was sufficient to do these replacements for the lattice with the
largest $a$, since the relative contribution of the boundaries to the
field variables residing on the bulk of lattice points among a given
gauge field configuration decreases with decreasing lattice spacing.
Therefore, the leading scaling violations can be regarded as being
purely $\Or(a^2)$ within our precision.

\begin{figure}[htb]
\begin{center}
\epsfig{file=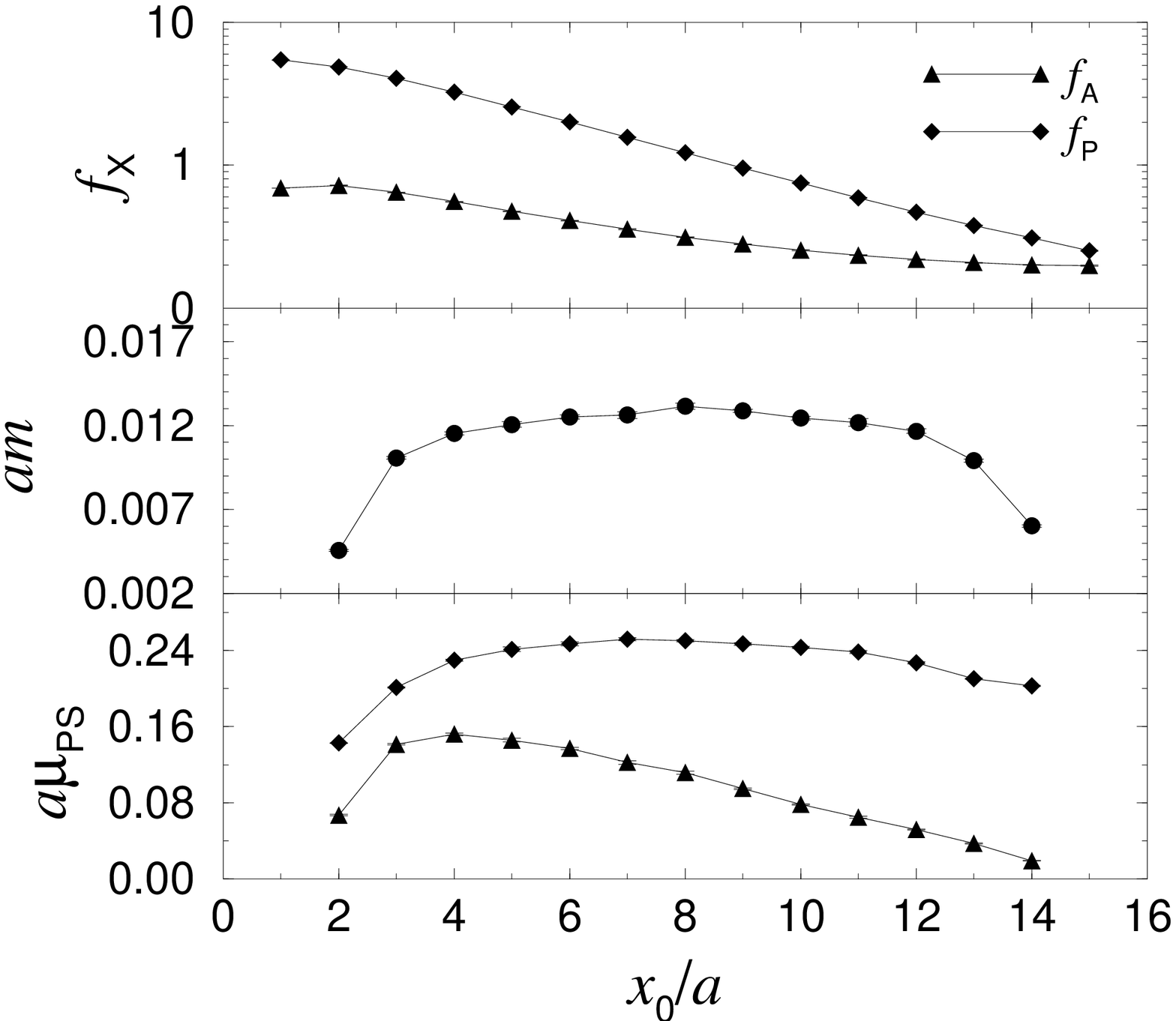,width=7.0cm}
\epsfig{file=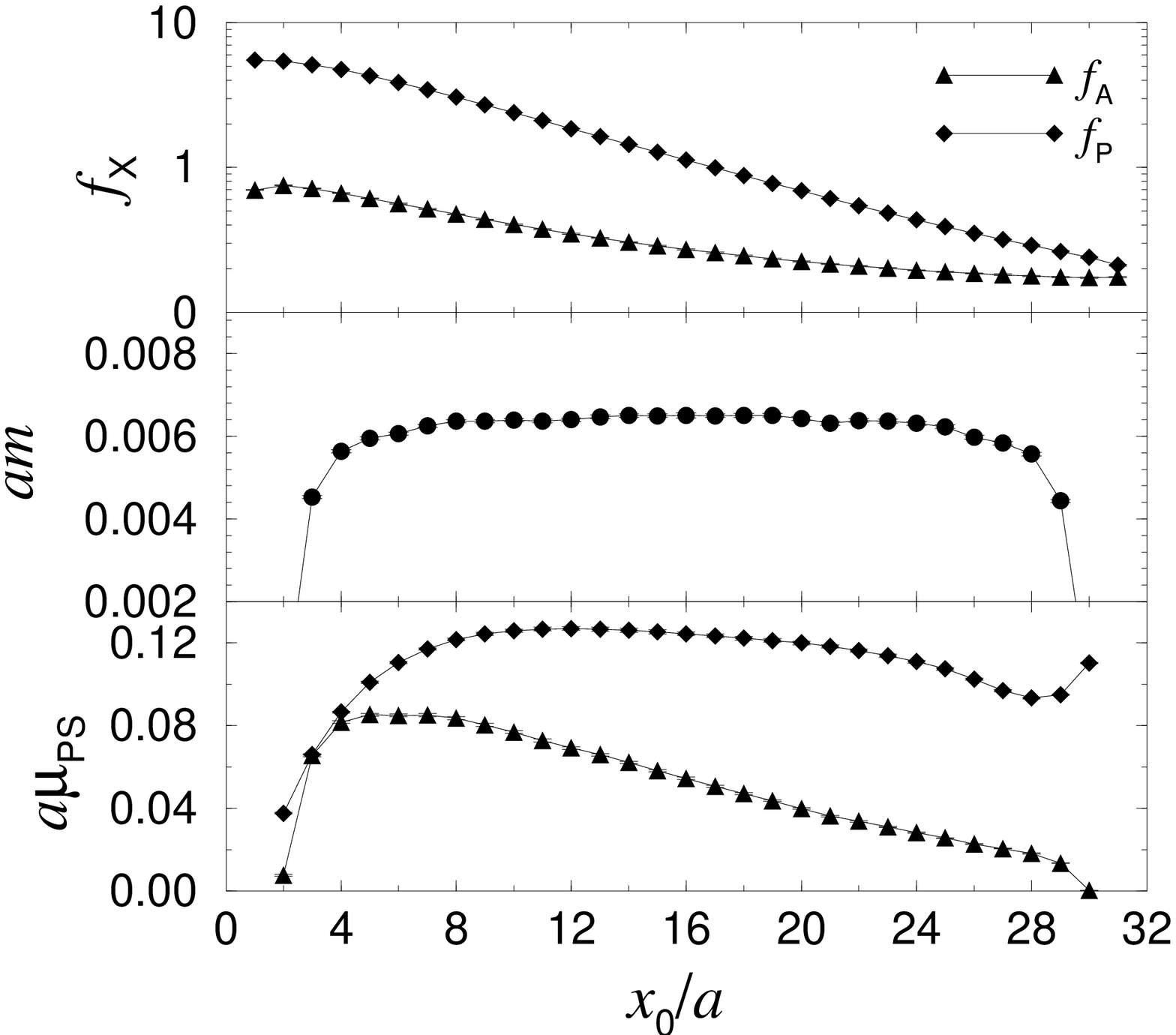,width=7.0cm}
\caption[t_param]{\label{CorrPlot} \sl
                  Correlation functions $\fa$ and $\fp$ (upper parts) and
                  $\Or(a)$ improved local masses extracted from them:  
                  PCAC current quark mass via eq.~(\ref{mcurr_x0})
                  (middle part) and `pseudoscalar meson masses' via
                  eq.~(\ref{mpi_x0}) and the same with $\fp$ substituted
                  by $\fai$ (lower parts).
                  The left figure depicts simulation point A ($L/a=8$)
                  and the right one point D ($L/a=16$), with $T/L=2$ in
                  both cases.
                  The solid lines are only meant to guide the eye, and the
                  statistical errors are smaller than the symbols.}
\end{center}
\end{figure}

In figure~\ref{CorrPlot} we first illustrate for simulation points A and D
the correlation functions $\fa$, $\fp$ and some of the quantities deduced
from them in the previous section in dependence of the time coordinate.
%
%
The correlators reflect that a good signal remains also at larger distances
in time.
The PCAC quark mass (\ref{mcurr_x0}) already exposes a plateau for the
rather moderate temporal extensions of the lattice, whereas the
`local pion mass' (\ref{mpi_x0}), and even more the logarithmic derivative
of the improved correlation of the axial current (\ref{cor_fa_i}), show
significant contributions from higher intermediate states in small volumes.
On physically large volumes, however, this pattern disappears:
plateaux develop around $x_0=T/2$, i.e.~the lightest excitations, which
coincide with the pseudoscalar meson mass when defined either by $\fai$ or
by $\fp$, govern the exponential decay of these functions.
This will be explicitly demonstrated elsewhere \cite{msbar:pap2}.
An analogous statement holds for the vector meson mass via the correlation
functions $\kvi$ and $\kt$.

The expectation values in the simulation points of table~\ref{ParTab} for
the observables eqs.~(\ref{mcurr_x0}), (\ref{mpi_and_fpi}) and
(\ref{mrho_and_frho}) are now summarized in table~\ref{ResTab1}.
Within all potential sources of errors to be incorporated in the analysis,
i.e.~the statistical one and those coming from $\za/\zp$, $\za$, $\zv$,
$L/r_0$ and $\mpi L$, the contributions caused by the uncertainties of the
renormalization factors $\zX$ ever dominate the combined errors quoted in
the second parentheses in the table.
Furthermore, any inherent small mismatch of the central values with the
renormalization conditions (\ref{cond_LCP}) on $\mpi L$ and $L/r_0$ in sets
B, C and D was corrected by a conservative estimation of the slopes
$\partial\cl{O}/\partial(a\mpi)$ and $\partial\cl{O}/\partial\beta$ with
$\cl{O}=\cl{O}(\mpi L,L/r_0,L/a)\in\{am,a\mrh,a\fpi,\etVb\}$, which enter
the identities for the required partial derivatives
\be
\frac{\partial\cl{O}}{\partial(\mpi L)}\simeq
\frac{a}{L}\,\frac{\partial\cl{O}}{\partial(a\mpi)}
\label{deriv_mpi}
\ee
\be
\frac{\partial\cl{O}}{\partial(L/r_0)}\simeq
\frac{a}{L}\,\frac{\partial\beta}{\partial(a/r_0)}\,
\frac{\partial\cl{O}}{\partial\beta}\simeq
\frac{r_0}{a}\,\frac{\partial\cl{O}}{\partial(L/a)}\,.
\label{deriv_beta}
\ee
They were numerically extracted in linear approximation from several
simulations in set A at neighbouring values of $\kappa$ and $\beta$,
where in the latter case they had to be combined with the derivative
$\partial\beta/\partial(a/r_0)$ to be read off from the
parametrization eq.~(\ref{r0_formula}).
The slopes obtained in this way were carried over to the finer lattices as
well, since for increasing lattice resolution ($\beta>6.0$) their
corrections are of $\Or(a)$ and, with respect to the other sources of
errors, can safely be ignored.
%
\begin{table}[htb]
\begin{center}
\begin{tabular}{|c|cccc|}
\hline
  set & $am(\frac{T}{2})$ & $a\mrh$ & $a\fpi$ & $\etVb$          \\
\hline
  A   & 0.0132(2)  & 0.3438(10) & 0.1248(5)(13) & 0.3283(19)(25) \\
  B   & 0.0083(1)  & 0.2688(15) & 0.1013(7)(12) & 0.3387(36)(40) \\
  C   & 0.0096(1)  & 0.2337(14) & 0.0851(7)(11) & 0.3289(40)(43) \\
  D   & 0.00651(6) & 0.1689(10) & 0.0650(5)(8)  & 0.3561(41)(44) \\
\hline
\end{tabular}
\caption[t_param]{\label{ResTab1} \sl 
                  Analysis results for the observables under consideration
                  in lattice units.
                  The first error is only the statistical one, the
                  second one (where given) includes in addition the
                  uncertainties from the renormalization constants.}
\end{center}
\end{table}
%

\section{Results}
\label{Res}
We pass to the final results.
After the procedure of matching the conditions characterizing the LCP
under study, one finds the numbers collected in table~\ref{ResTab2};
note that according to (\ref{mpcac}) the bare PCAC quark mass
$m(\frac{T}{2})$ translates through multiplication with $\za/\zp$ into
the renormalized quantity $\mbar$.
%
\begin{table}[htb]
\begin{center}
\begin{tabular}{|c|cccc|}
\hline
  set & $\mbar r_0$ & $\mrh r_0$ & $\fpi r_0$ & $\etVb$    \\
\hline
  A   & 0.1069(50)  & 1.846(13)  & 0.6701(82) & 0.3283(46) \\
  B   & 0.1029(35)  & 1.839(15)  & 0.684(11)  & 0.3296(72) \\
  C   & 0.1045(36)  & 1.848(18)  & 0.681(13)  & 0.3360(95) \\
  D   & 0.1103(35)  & 1.860(18)  & 0.699(12)  & 0.3457(75) \\
\hline
\end{tabular}
\caption[t_param]{\label{ResTab2} \sl
                  Dimensionless results for the quantities of
                  table~\ref{ResTab1} with total errors after they have
                  been corrected to fulfill exactly the simultaneous
                  renormalization conditions (\ref{cond_LCP}) as described
                  in the text (subsection~\ref{Meth}).
                  The errors on $r_0/a$ quoted in \cite{pot:r0_ALPHA}
                  have been taken into account as well.}
\end{center}
\end{table}
%
Now we are prepared to perform extrapolations of these data to the
continuum limit, assuming convergence with a rate proportional to $a^2$.

\begin{figure}[htb]
\begin{center}
\epsfig{file=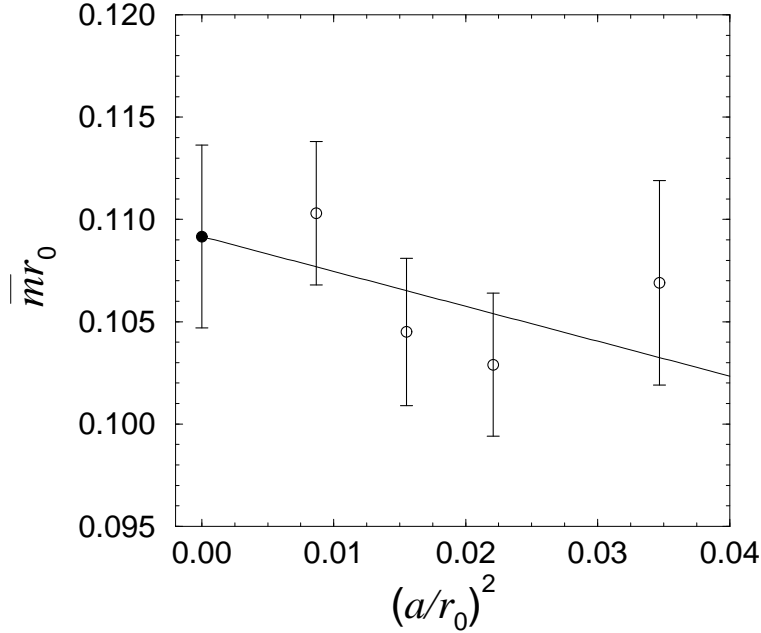,width=10.0cm}
\caption[t_param]{\label{MbarPlot} \sl
                  Scaling behaviour of the renormalized PCAC current quark
                  mass in units of $r_0$. 
                  An intermediate volume with SF boundary conditions is
                  considered.
                  The renormalization constant $\zp$ refers to a scale
                  of $L=1.436\,r_0$.                   
                  Non-perturbative $\Or(a)$ improvement allows to
                  extrapolate linearly with $(a/r_0)^2\rightarrow 0$
                  to the continuum limit.}
\end{center}
\end{figure}

\begin{figure}[htb]
\begin{center}
\epsfig{file=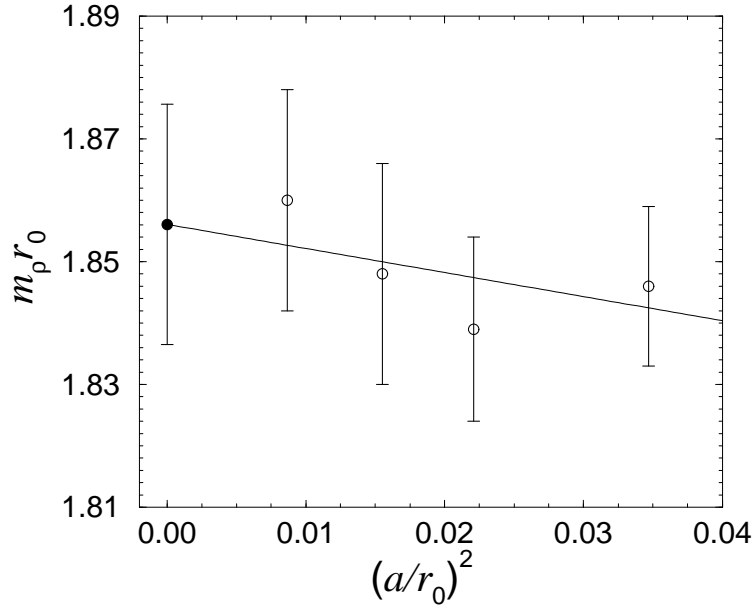,width=10.0cm}
\caption[t_param]{\label{MrhoPlot} \sl 
                  Continuum limit extrapolation as in figure~\ref{MbarPlot}
                  but for the `rho meson mass' in intermediate volume.}
\end{center}
\end{figure}

\begin{figure}[htb]
\begin{center}
\epsfig{file=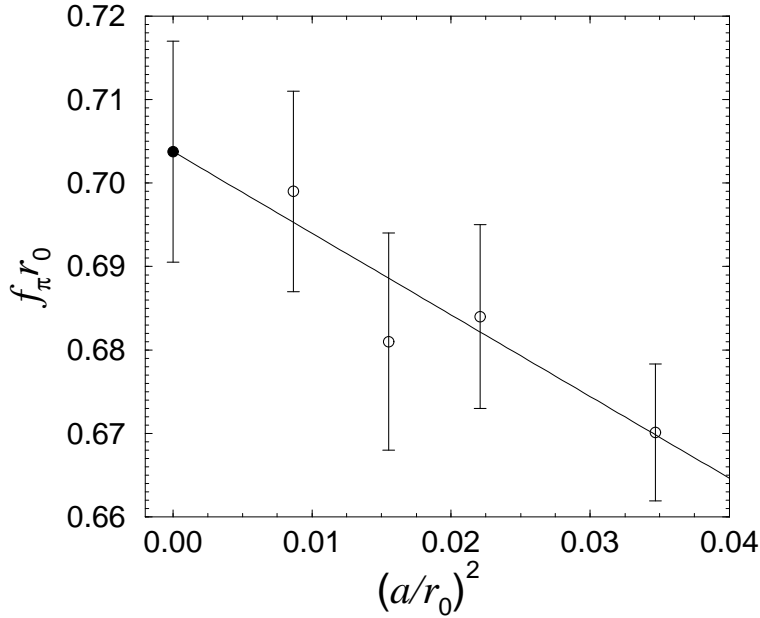,width=10.0cm}
\caption[t_param]{\label{FpiPlot} \sl 
                  The same as in figure~\ref{MrhoPlot} but for the
                  `pion decay constant'.}
\end{center}
\end{figure}

\begin{figure}[htb]
\begin{center}
\epsfig{file=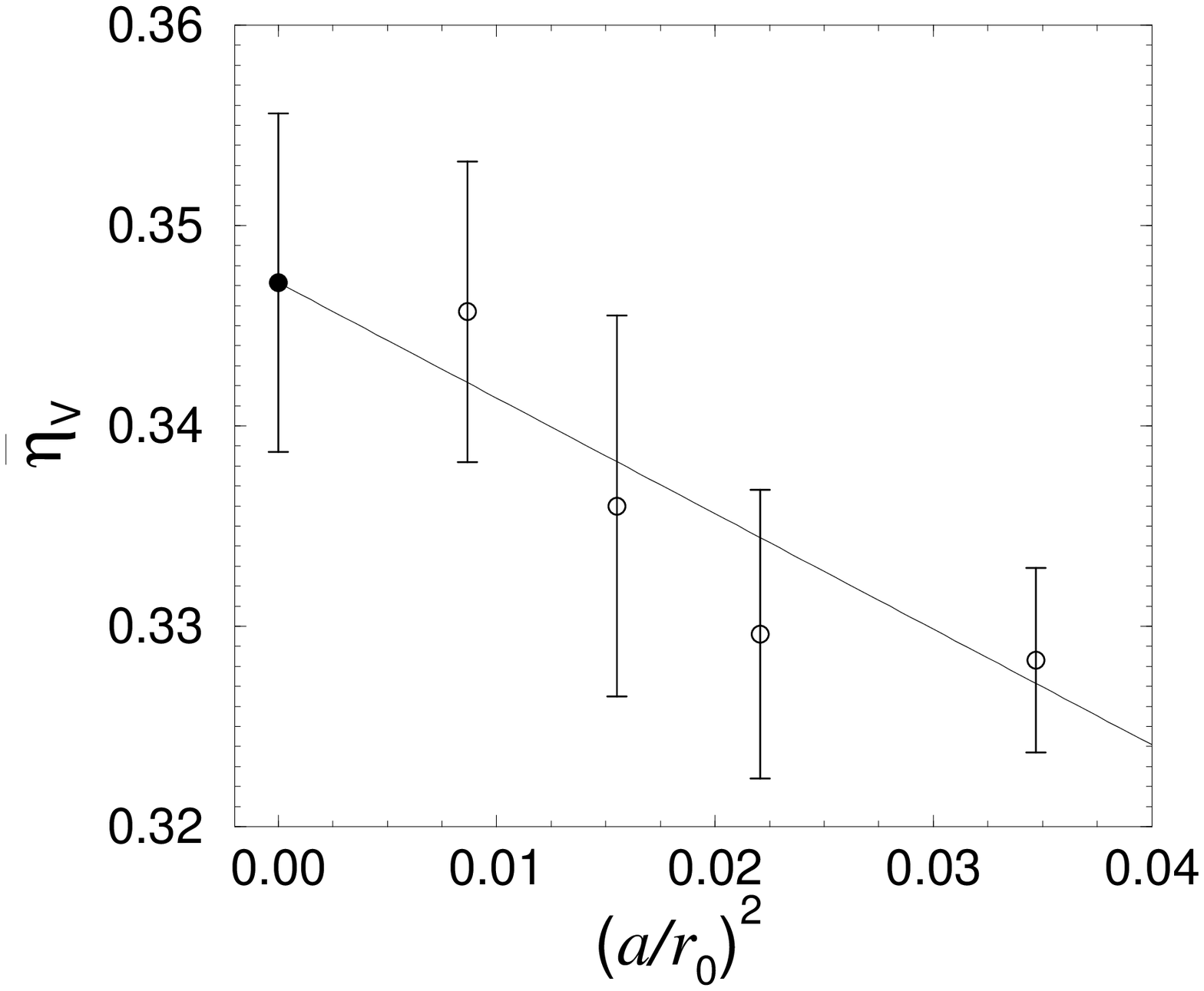,width=10.0cm}
\caption[t_param]{\label{FrhoPlot} \sl
                  Again as in figure~\ref{MrhoPlot} but now for the
                  observable $\etVb$, which is composed of a
                  (renormalized and improved) correlation function
                  involving the vector current.}
\end{center}
\end{figure}

The fits are displayed in figures~\ref{MbarPlot} -- \ref{FrhoPlot} and
exemplify the scaling behaviour on the course from simulation points
A to D.
%
%
%
%
%
%
%
%
One evidently observes the leading corrections to the continuum to be
compatible with $\Or(a^2)$.
Moreover, it can be inferred from table~\ref{CLimTab} that the differences
of the continuum limits from the values at $\beta=6.0$
($a\simeq 0.1\,\,\mbox{fm}$) are around or even below 5 \% in the improved
theory.
%
\begin{table}[htb]
\begin{center}
\begin{tabular}{|c|cccc|}
\hline
  case & $\mbar r_0$ & $\mrh r_0$ & $\fpi r_0$ & $\etVb$    \\
\hline
  fully improved   
       & 0.1092(45)  & 1.856(20)  & 0.704(13)  & 0.3472(84) \\
       & 2 \%        & $<$ 1 \%   & 5 \%       & 6 \%       \\
\hline
  $\ca=0$ and $\cv=0$ 
       & 0.1069(46)  & 1.856(20)  & 0.698(13)  & 0.3439(85) \\
       & 17 \%       & $<$ 1 \%   & 1 \%       & 2 \%       \\
\hline
\end{tabular}
\caption[t_param]{\label{CLimTab} \sl
                  Continuum limits and their percentage deviations from
                  $\beta=6.0$ ($a\simeq 0.1\,\,\mbox{fm}$).
                  The lower numbers belong to a data evaluation, where the
                  improvement coefficients $\ca$ and $\cv$ have been
                  artificially taken to vanish.}
\end{center}
\end{table}
%
These appear to be partly smaller than it was to be expected on basis of
the experiences reported previously in
refs.~\cite{impr:QCDSF,lat97:hartmut}.

At this point we have to add the remark that the results in the vector
channel had to be revised compared to those listed in
\cite{lat98:jochen}.
Due to some incorrect normalization of the correlation functions $\kv$ and
$\kt$ during an earlier data analysis, we erroneously observed a quite
steep slope when carrying out the $(a/r_0)^2\rightarrow 0$ fit for the
ratio $\etVb$ defined in eq.~(\ref{mrho_and_frho}).
Looking at the final numbers now, it does no longer stand in contradiction
to the findings in the pseudoscalar channel.
By contrast, since the scaling violations of $\etVb$ are only slightly
larger than for $\fpi r_0$, we interpret this as a further compelling and
satisfactory indication for the effectiveness of non-perturbative $\Or(a)$
improvement.

In ref.~\cite{lat97:hartmut} the suspicion was raised that scaling looks
even slightly better, if the perturbative estimates for the improvement
coefficients $\ca$ and $\cv$ are used.
This issue deserves some comments here.
In order to address the sensitivity of the analysis to the improvement
terms in the pseudoscalar and vector channel proportional to $\ca$ and
$\cv$, we just evaluated our data with setting these coefficients
arbitrarily to zero.
Then, of course, the theory is only partially $\Or(a)$ improved, and a
residual contamination with uncancelled $\Or(a)$ contributions has still
to be expected.
The surprising outcome is that upon omitting those terms, two quantities
($\fpi r_0,\etVb$) have somewhat smaller $a$--effects in total magnitude.
At the same time, however, the renormalized PCAC quark mass $\mbar r_0$
gets much larger ones.
Nevertheless, as seen from table~\ref{CLimTab}, the continuum limits of
both data sets agree within errors.
Such a result might suggest that the \emph{qualitative} scaling behaviour
of our quantities (apart from $\mbar r_0$) is only marginally influenced 
by the definite choice of $\cX$, X=A,V, still meeting the condition of a
dominant $a^2$--behaviour.
On the other hand, we observed a tendency in the $\ca=\cv=0$ data points
to disperse around the straight line fits to the continuum, which hints
at some remnant of admixture of $\Or(a)$ discretization errors;
hence a scaling violating term $\propto a^2$ in leading order rather
seems to be ruled out in that case.
This is particularly pronounced for $\mbar r_0$, where the correction to
the continuum limit grows distinctly
(from 2 \% to 17 \% in table~\ref{CLimTab}) if the improvement of the
axial quark current is switched off. 
Opposed to that, in the fully improved case including the $\Or(a)$
correction terms, the required continuum limit extrapolations are
generically not critical.
In conclusion --- and as a further convincing argument for the use of its
non-perturbative values --- this finally supports the physical insight
that $\ca$ and $\cv$ are essentially relevant for chiral symmetry
restoration at finite cutoff.

\section{Discussion and outlook}
\label{DiscOut}
In confirmation of similar investigations
\cite{impr:QCDSF_lett,impr:QCDSF,lat97:hartmut,impr:SCRI}, $\Or(a)$
improvement implies a substantial reduction of scaling violations.
Our numerical simulations of renormalized correlation functions in
intermediate physical volume within the Schr\"odinger functional give
clear evidence for an overall behaviour completely consistent with being
linear in $a^2$ at $a\le 0.1\,\,\mbox{fm}$, for all quantities under
consideration.
Changing $a$ by a factor two yields very stable fits and honest continuum
limit extrapolations.
Actually, the residual $\Or(a^2)$ cutoff effects at
$a\simeq 0.1\,\,\mbox{fm}$ stay around ($\fpi r_0$ and $\etVb$) or 
significantly below 5 \% ($\mbar r_0$ and $\mrh r_0$).

To quantify directly the influence of the improvement coefficients $\ca$
and $\cv$ on the scaling behaviour, we examined also the partially
improved case $\ca=\cv=0$.
In this case $\fpi r_0$ and $\etVb$ show an even weaker dependence on the 
lattice spacing, \emph{but now one finds $\sim$ 17 \% lattice spacing
effects} at $a\simeq 0.1\,\,\mbox{fm}$ in the renormalized current quark
mass $\mbar r_0$.
Additionally, as outlined in section~\ref{Res}, the functional form of the
leading $a$--effects then appears no longer compatible with $a^2$ alone;
furthermore, chiral Ward identities are badly violated for $\ca=\cv=0$
at $\Or(a)$ level \cite{impr:pap3,lat97:marco}.
Thus there is in general no choice for the improvement coefficients $\ca$
and $\cv$, which diminishes the size of $\Or(a^2)$ lattice artifacts
simultaneously for \emph{all} relations and observables below a level of
5 \%.
However, one should not feel tempted to judge this fact as a kind of
principal conflict with the improvement programme itself, since the
criterion of small $\Or(a^2)$ corrections has only been touched when
selecting a definite set of kinematical variables to formulate the
respective improvement conditions within the Schr\"odinger functional.

To summarize, the scaling tests in hand illustrate that also in the
$\Or(a)$ improved theory the remaining $\Or(a^2)$ discretization errors
have to be assessed --- and consequently can be extrapolated away
reliably --- by varying the lattice spacing.

An extension of the present study to physically large volumes, where
hadronic masses and matrix elements can be computed, is in progress
\cite{msbar:pap2}.
\subsection*{Acknowledgements}
This work is part of the ALPHA collaboration research programme.
We thank DESY for allocating computer time on the APE-Quadrics computers
at DESY Zeuthen to this project and the staff of the computer centre at
Zeuthen for their support.
I am grateful to Rainer Sommer for numerous discussions, useful 
suggestions and a critical reading of the manuscript.
I also thank Hartmut Wittig and Andreas Hoferichter for discussions.

\bibliography{lattice_ALPHA}     
\bibliographystyle{h-elsevier}   
\end{document}